\documentclass[aps,pra,showpacs,twoside,twocolumn,10pt]{revtex4-1}
\usepackage[colorlinks=true, citecolor=red, urlcolor=blue ]{hyperref}
\usepackage{epsfig,newlfont,amssymb,amsfonts,amsmath,bm,subfigure,palatino,mathtools,amsthm,braket,soul,enumitem,color,graphics,graphicx,times,physics}

\usepackage[normalem]{ulem}
\usepackage{xcolor}

\begin{document}
\title{
Quantum Battery with Ultracold Atoms: Bosons vs. Fermions
}

\author{Tanoy Kanti Konar,  Leela Ganesh Chandra Lakkaraju, Srijon Ghosh, Aditi Sen(De)}
\affiliation{Harish-Chandra Research Institute, A CI of Homi Bhabha National Institute, Chhatnag Road, Jhunsi, Prayagraj 211 019, India}

\begin{abstract}
We design a quantum battery made up of bosons or fermions in an ultracold-atom setup, described by
\emph{Fermi-Hubbard} and \emph{Bose-Hubbard} models, respectively. We compare the performance of bosons and fermions to determine which can function as a quantum battery more effectively given a particular on-site interaction and initial state temperature. The performance of a quantum battery is quantified by the maximum energy stored per unit time over the evolution under an on-site charging Hamiltonian. We report that when the initial battery state is in the ground state, fermions outperform bosons in a certain configuration over a large range of on-site interactions which are shown analytically for a smaller number of lattice sites and numerically for a considerable number of sites. Bosons take the lead when the temperature is comparatively high in the initial state for a longer range of on-site interaction. We study a number of up and down fermions as well as the number of bosons per site to find the optimal filling factor for maximizing the average power of the battery. We also introduce disorder in both on-site and hopping parameters and demonstrate that the maximum average power is robust against impurities. Moreover, we identify a range of tuning parameters in the fermionic and bosonic systems where the
disorder-enhanced power is observed.

\end{abstract}

\maketitle

\section{Introduction}

In recent years,  tremendous efforts have been devoted to decorate the avenue of quantum technologies which includes the advancement of miniaturized quantum devices ~\cite{popescu10}, that are indispensable in various practical purposes. Such quantum gadgets are shown to outperform the existing classical ones in different sectors ranging from metrology \cite{Giovannetti2011},  cryptography \cite{gisinrmp}, cybersecurity, and to data analysis and  computing \cite{Yang21}. The development of smaller, and more effective devices naturally leads to the realm of quantum mechanics. In this respect, microscopic thermodynamic devices are also shown to provide a remarkable precision in thermometry \cite{joulain16}, thereby contributing to the field of quantum thermodynamics \cite{gemmer2004,sai2016}. To explore and model the quantum thermal machines such as quantum refrigerators\cite{popescu10,chiara2020,mitchison2015,correa2013,brask2015,das2019}, quantum batteries\cite{Alicki,Zhang2010,zhao2021,Binder2015,campaioli2017,srijon2020,andolina2020,santos2019,Dou_2020, srijon21,Batteryreview},  modified definitions of work, heat, and entropy are introduced that can take into account the effects of quantumness in the system.

 The behavior of traditional chemical batteries that can store energy is purely classical in nature, and hence  cannot be used in quantum mechanical apparatus. With this requirement,  Alicki and Fannes first proposed the concept of quantum battery  (QB) \cite{Alicki}, a $d$-dimensional quantum mechanical system composed of $N$ non-interacting  subsystems which are able to store energy for future use and  can  efficiently be  charged by global entangling operations. After the initial proposal, several interesting works were reported \cite{Batteryreview} which include quantum batteries with Dicke state \cite{andolina2017, andolina2019}, the role of entanglement-production  in the process of work-extraction \cite{Hovhannisyan2013, Binder2015}, non-local charging having an extensive advantage in power storage \cite{dario21}, and the effects of decoherence on quantum batteries \cite{barra2019,liu2019, Giovannetti2019, quach2020,tejero2021, srijon21}. Specifically,  when  the battery is in contact with  the environment, both in presence of  Markovian and non-Markovian noises,  it was reported that  non-Markovian noise can sometimes help to extract high amount of work depending upon system parameters \cite{Giovannetti2019, srijon21}.   
 On the other hand, interacting spin systems composed of spin-$s$ particles can also be used to design QBs which can be charged via local magnetic field \cite{Modispinchain, srijon2020, srijon2021}. In a similar spirit, the nearest-neighbor hopping interaction of a spin chain acted as a battery and coupled with a cavity mode is shown to enhance the capability of storing energy in the system \cite{zhao2021}. 
More importantly, quantum batteries are shown to be realized in different platforms like solid-state systems where each of the two-level systems are either enclosed in a single cavity or the ensembles of two-level systems is in a single cavity \cite{andolina2017, recentexperiment}, and  superconducting circuits which can be charged by using external magnetic field  \cite{hu2021optimal}.  \\

In this work, we propose to design a quantum battery with a one dimensional $\emph{Hubbard model}$, realizable via cold atoms in an optical lattice, where the lattice is filled up with either fermions or bosons, well-described by the $\emph{Fermi-Hubbard}$ (FH) and  the $\emph{Bose-Hubbard}$ (BH) \cite{aditireview, blochRMP08, esslinger2010, Dutta_2015} models respectively (see Fig. \ref{fig:schematic}). Specifically, the initial state of the battery is prepared as the ground or the canonical equilibrium states of the FH and the BH models while  the charging of the battery can take place by tuning the strength of  the on-site intraatomic  interactions. 
It is important to stress here that in all the aforementioned proposals of QBs,  the subsystems are distinguishable as their positions are fixed in space while in the current proposal,  the particles can hop from one lattice site to others and as a consequence become indistinguishable within the lattice system. 
We also know that both the models possess rich phase diagrams having phases like mott insulator,  superfluid, superconducting, Fermi liquid  \cite{fisher1989, carr2005,aditireview},   and  density-wave,  Haldane insulator phases  in the extended BH model \cite{Freericks1994,Rossini2012} and hence such a study may connect physical properties of the bosonic and fermionic systems with quantum
thermodynamics.

A comparative study carried out  between  the FH and the BH  models reveals that the fermionic batteries with more than two lattice sites can store a higher amount of extractable power output than those of bosonic systems provided the repulsive or  attractive on-site interactions  are  suitably tuned by varying the scattering lengths and the initial state of the battery is at the zero temperature with half-filling. The hierarchy gets reversed, i.e., the batteries made up of BH models demonstrate advantage over the FH ones, when the initial state is prepared at a finite and high temperature.
We also illustrate that apart from the ratio between the intraatmoic on-site and the interatomic hopping interactions, the patterns of the power output also depend on the even or odd lattice sites in both  models. 
For a fixed lattice site, we optimize the maximum average power output   over configurations allowed for fermions  and bosons where in the latter case, we also fix the particles per site and observe that the optimized power decreases (increases) with the increase of lattice sites (the increase of  the particles per site) for fermions (bosons). 


With the significant advancement in  experiments having different physical substrates,  the disordered quantum systems \cite{fort2005,fallani2007,white2009,Shapiro2012,ahufinger2005} are of great interest to study since it is almost impossible to prepare a system avoiding the impurities in laboratories. In particular, although the high level of control to generate and manipulate  ultracold bosonic, or fermionic gases, or their mixtures have been achieved experimentally, there is always a possibility to include disorder  during the transfer of the samples to optical lattices.  Disorder in the hopping parameter of the Hubbard Hamiltonian can also be realized by modulating the applied electric field of the laser or by doping impurities in the system.  Impurities can also appear by aid of additional lattices or by modulating magnetic fields in the system \cite{aditireview, Damski03}. Interestingly, cold atomic systems turn out to be one of the experimental-friendly platforms where disordered systems can be realized and engineered. 

Although, intuitively, disorder  detrimentally affects the characteristics of the quantum systems and hence the performance, it was shown to be not true \cite{aharony1978, minchau1985, wehr2006, abanin2007, hide2009, Niederberger2010, prabhu2011,  sadhukhan2016, srijon2020, srijon2021}, i.e.,  certain features of the quantum system are found to get enhanced even in presence of impurities.
Moreover, disordered systems show a lot of counter-intuitive phenomena which include Anderson localization \cite{Anderson58}, many-body localization which pinpoints the distinction between thermalization and localized phase  \cite{blochrmp, bardarson2012, pal2010, canovi2011}, and high-temperature superconductivity \cite{rosenstein2010}  to name a few.  
In this respect, we show that the quenched averaged power outputs  are robust against  random hopping and random  on-site interactions in  both FH and  BH models. In case of disorder introduced in hopping, we report that there is a regime of the hopping strength in which both bosonic and fermionic disordered systems can produce higher maximal power than that of the ordered ones which we refer to as the \emph{disorder enhanced power}. Such increment in power can be seen due to the monotonically increasing nature of power in batteries with ordered Hamiltonian. The  randomness in hopping and on-site interactions are chosen from Gaussian as well as uniform distributions with a fixed mean and standard deviations and both types of randomness can be realized in cold atomic setup.

The paper is organized in the following manner. The design of the quantum battery based on Hubbard models and their charging processes are introduced in Sec. \ref{sec:model}. In the next section (Sec. \ref{sec:analytical}), the performance of the QB and the comparative studies between bosonic and fermionic systems are carried out. In Sec. \ref{sec:fillingtemp}, the effects of the filling factor and the temperature of the initial state are investigated while the disordered BH and FH models are considered as batteries in Sec. \ref{sec:robust}. Finally, the concluding remarks are discussed in Sec. \ref{sec:conclu}. 

\section{ Modelling of Quantum battery using Hubbard Hamiltonians}
\label{sec:model}


\begin{figure}
     \centering
     \includegraphics[scale=0.35]{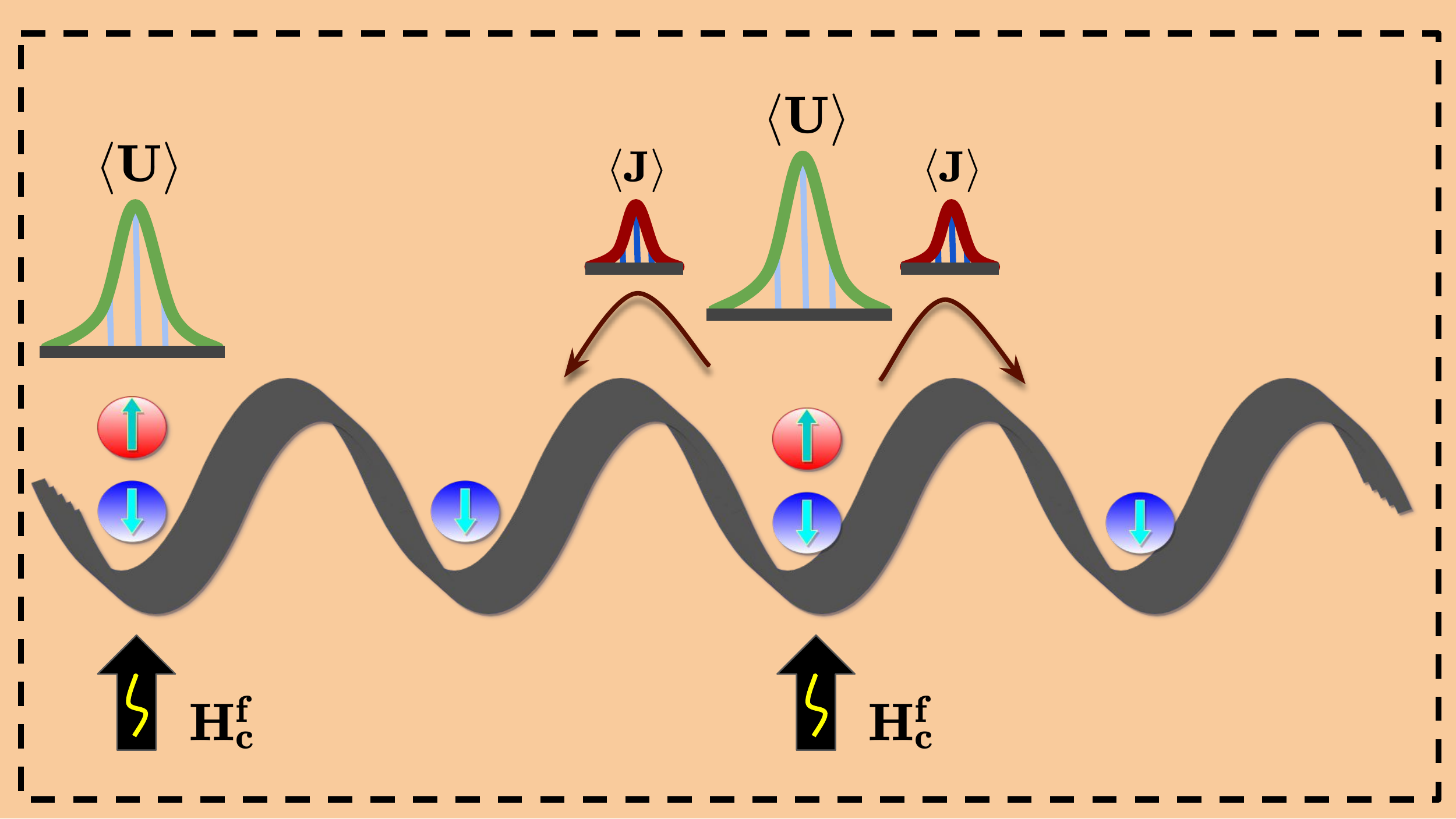}
     \caption{ (Color Online.) Schematic representation of a quantum battery based on one-dimensional Fermi-Hubbard Hamiltonian with and without disorder with \(L\) lattice sites. \(H_c^f\) in Eq. (\ref{eq:charging}) is used to charge the battery. As shown in the paper, similar modelling of QB is also possible with Bose-Hubbard system. Notice that the charging part solely depends on the on-site interaction term, $U^{\mu}_i$ (\(\mu =b, f\)), which is  nonvanishing for a particular lattice site only when  atleast two particles are  present at each lattice. However,  there are hopping terms   for any two sites  which we have not shown for all sites  to make the schematics less cumbersome. We also report the robustness observed in the performance of the QB against different kinds of disorder, present in the on-site as well as in the hopping strength.
     }
     \label{fig:schematic}
 \end{figure}

We model a quantum battery as the  one-dimensional Hubbard Hamiltonian (see Fig. \ref{fig:schematic} for schematic representation) with $L$ lattice sites filled with fermions or bosons, known as \emph{Fermi-Hubbard} and \emph{Bose-Hubbard}  models respectively which can be engineered in the laboratory with cold atoms in optical lattices \cite{blochRMP08, aditireview}. Such a study also identifies the regime where  fermionic systems show a better performance as a QB than that of the bosonic ones and vice-versa. 
 
 \emph{Model of the battery.} The initial state of a quantum battery \cite{Alicki, Modispinchain, srijon2020}  is taken as  the ground state or the  canonical equilibrium state i,e.,  $\rho_{th} = \frac{e^{-\beta' H_{B}^{\mu}}}{Z}$,  (\(\mu = f, b\)) of the Fermi-Hubbard or Bose-Hubbard Hamiltonian, $H_{B}^{\mu}$, where the superscript represents fermionic or bosonic systems. Here $\beta' = \frac{1}{k_{B} T}$ with $k_{B}$ being the Boltzmann constant, $T$ being the absolute temperature and  the partition function, $Z = \Tr[e^{-\beta' H_{B}^{\mu}}]$.  When  the lattice sites are occupied with fermions,   \emph{Fermi-Hubbard Hamiltonian} can be represented as
\begin{equation}
    H_{B}^{f}=-\sum_{<ij>,\sigma} J^{f}_{ij} c_{i\sigma}^{\dagger}c_{j\sigma} +h.c +\sum_{i} U^{f}_{i} n_{i\uparrow}n_{i\downarrow}.
    \label{Hamiltonian_fermi}
\end{equation}
Here $J^{f}_{ij}$ is the hopping strength between the sites, $i$ and $j$, where $\langle ij \rangle$ indicates  that the nearest-neighbor hopping is only allowed, and $U^{f}_{i}$ denotes the on-site interaction at the site i of the lattice, occupied by fermions, which can be repulsive as well as attractive. 
$c_{i\sigma}$ ($c_{i\sigma}^{\dagger}$) is fermionic annihilation (creation) operator obeying the canonical anti-commutation relations, $\{c_{i\sigma},c_{j\sigma '}^{\dagger}\}=\delta_{ij}\delta_{\sigma\sigma '}$, $\{c_{i\sigma},c_{j\sigma '}\}=0$ and $\{c_{i\sigma}^{\dagger},c_{j\sigma '}^{\dagger}\}=0$, and  $n_{i \sigma} = c_{i \sigma}^{\dagger}c_{i \sigma}$ is the number operator on the site $i$ having spin $\sigma$.
 
Instead of  fermions, when the lattice sites are filled with bosons, the \emph{Bose-Hubbard Hamiltonian} reads as
\begin{equation}
    H_{B}^{b}=-\sum_{<ij>} J^{b}_{ij}b_{i}^{\dagger}b_{j} +h.c +\sum_{i} \frac{U^{b}_{i}}{2} n_{i}(n_{i}-1),
\end{equation}
where $J^{b}_{ij}$ and $U^{b}_{i}$ are the  hopping strength from site $i$ to $j$ and the on-site interaction strength at the $i$-th site respectively, and $b_{i}$ ($b_{i}^{\dagger}$) is bosonic annihilation (creation) operator following  the standard canonical commutation relations for bosons.  For both the models, we consider the open boundary condition (OBC). Moreover, from now on, site-independent parameters will be denoted  without the subscripts, $i$ and $j$, thereby representing the ordered models.

 \emph{Charging.}  In order to charge the system, we construct the charging Hamiltonian for fermions and bosons respectively as
\begin{equation}
    H_{c}^{f}=U_{c}^{f}\sum_{i} n_{i\uparrow}n_{i\downarrow}, \,\,
\mbox{and}\,\, 
  H_{c}^{b}=\frac{U_c^{b}}{2}\sum_{i} n_{i}(n_{i}-1).
  \label{eq:charging}
\end{equation}
Here $U^{\mu}_{c}$ is the charging strength and   in general, $U^{\mu}_{c} \ne U^{\mu}$. The reason for choosing such a form of the charging Hamiltonian is three-fold.  Firstly, we choose the charging Hamiltonian in such a way that the evolution is non-trivial. Precisely, the charging Hamiltonian should not commute with the battery Hamiltonian, otherwise  the system  cannot evolve. To ensure this, we put $J^{\mu} = 0$ in the original Hamiltonian, and construct the charging Hamiltonian even with different strength of charging, denoted by, $U_{c}^{\mu}$, which is, in general, not equal to $U^{\mu}$ ($\mu =$ f or b) of the battery Hamiltonian, \(H_B^{\mu}\).
Secondly, a charging Hamiltonian with $J \ne 0$ and $U = 0$ represents a global charging  which we do not wish to analyze,
since the extra increment in power can happen due to the global charging in the system. Finally, from the experimental point of view, adjusting $U$ is  much easier  than controlling the other parameters of the system. In an ultracold atom setup, the intensity can be tuned just by a ``control nob" as demonstrated in Refs. \cite{moeckel2008, chen2011, bertini2017}. E.g., the ratio of  $U/J$ can be tuned  by the laser intensity which  regulates the lattice potential depth, $s$. Tuning the corresponding $s$ is the way to traverse across the quantum phase transition, i.e., between Mott and superfluid phases. In the design discussed above,  the battery is to be charged with the Bose- (Fermi-) Hubbard model having $J=0$, i.e., to be quenched  in the Mott-insulator phase. Specifically, when  $s > 13 E_r $ \cite{chen2011}, the system is in the Mott-insulator phase, where $E_r = h / 8 m d^{2} $, with $m$ being the atomic mass, $d = 406nm$ being the lattice spacing, and $h $ being the Planck's constant. The value of $s$ can be dynamically modulated to perform the quench without having any 
restriction on the time scale for the quench, and hence one can expect that the  charging of the battery can be achieved in a similar experiment.


\emph{Quantifying performance.} By employing unitary operations, $\mathcal{U}_{c} = \exp(-i H_{c}^{\mu} t)$, such that $\rho(t) = \mathcal{U}_{c} \rho(0) \mathcal{U}_{c}^{\dagger}$, with \(\rho(0)\) being the initial  state of the QB, 
the total amount of energy that can be stored as well as extracted  from the QB  (the work output) at time $t$ reads as
\begin{equation}
    W^{\mu} (t) = \Tr \left [H_{B}^\mu \rho(t)\right] - \Tr\left [H_{B}^{\mu} \rho(0)\right],
    \label{eqn.1}
 \end{equation}
where the first and the second terms in Eq. (\ref{eqn.1}) are the final and initial energies of the system respectively. Notice that the maximum amount of extractable work from the quantum battery in terms of ergotropy coincides with the above equation in the case of a reversible unitary process. This is due to the fact that for entropy-preserving unitary operations, the maximum amount of extractable work is equal to ergotropy \cite{hulpke2006,bera2019}. Note, moreover, that the stored energy, in general, do not coincide with the extractable work (ergotropy), when the battery Hamiltonian is in contact with the environment \cite{barra2019,liu2019, Giovannetti2019, quach2020,tejero2021, srijon2021}.  

The maximum average power output from the battery at time $t$, is quantified as
\begin{equation}
    P_{\max}^{\mu} = \max_{t} \frac{W^{\mu}(t)}{t}.
    \label{eqn.2}
\end{equation}
Throughout the paper, we will use  \(P_{\max}^{\mu}\) as the figure of merit for determining the performance of the QB. Notice also that \(P_{\max}^{\mu}=0\), when the hopping term of the QB vanishes.  
It is important to note that the charging Hamiltonian is turned on when \(t>0\) and
hence  the energy cost of turning on and off  the charging Hamiltonian is typically not  considered during the computation of power.



\emph{Scaling.} By increasing the values of $U_{i}^{\mu}$, and \(J_{i}^{\mu}\)  ($\mu = f, b$), of \(H_B^\mu\), it is possible to store more and more extractable power output from the quantum battery which makes the analysis trivial. To avoid such an undesirable situation, we normalize the Hamiltonian in such a way that its spectrum is bounded in $[-1,1]$, irrespective of any system parameters including the system size. In addition, since we compare between two different models, namely FH and BH, it is necessary to consider them in an equal footing from the perspective of energies, which can also be taken care by the normalization. Hence, the normalized Hamiltonians reads as
\begin{equation}
 \frac{1}{E_{max}-E_{min}}[2H_B^{\mu}-(E_{max}+E_{min})\mathbb{I}]\rightarrow H_B^{\mu},   
\end{equation}
where $E_{max}$ and $E_{min}$ are the maximum and the minimum eigenvalues of $H_{B}^{\mu}$. Here, since the evolution is unitary, the excitation spectrum is also conserved in both the cases, which also ensures that the comparison is fair.

\section{Performance of  QB for arbitrary number of lattice sites: comparing Bosons with fermions }
\label{sec:analytical}

Let us now concentrate on a hierarchy among QBs based on BH and FH models according to their performance.  We start with two lattice sites and then investigate the trends of the power output for arbitrary lattice sites. In this section, the number of particles is same as the number of lattice sites. 

\subsubsection*{Two-lattice sites: Equivalence between bosonic and fermionic systems }
First consider a  scenario when two particles occupy a lattice having two sites. In this situation,  the  work  output can be found analytically both for bosons and fermions and their relation is as follows. 

\textbf{Proposition 1.}
\emph{The average work outputs for BH and FH models coincide for a lattice with two sites occupied by two particles if the values of on-site interactions, hopping and charging strengths are identical, and the initial state of the battery is prepared as the ground state of the Hamiltonian.} 
\begin{proof}
The two-sites Fermi-Hubbard model occupied with two fermions has four basis states. Generically, the Fock state bases are defined as $|x_1 y_2\rangle_{\uparrow} |z_1 w_2\rangle_{\downarrow}$. Here
\(\{x_1, y_2, z_1, w_2\} \in (0,1)\), where \(0\) denotes the situation when the lattice site is not occupied by fermions while \(1\) is when the fermion occupies the lattice site and subscripts denote the lattice sites which we drop now on and we will use only the binary method to indicate the entire configuration. In this basis, the normalized Hamiltonian reads as 
\begin{equation*}
H_{B}^{f} = \frac{1}{\sqrt{{16J^f}^2+{U^f}^2}}\begin{bmatrix}
U^{f} & -2J^{f} & -2J^{f} & 0\\
-2J^{f} & -U^{f} & 0 & -2J^{f}\\
-2J^{f} & 0 & -U^{f} & -2J^{f}\\
0 & -2J^{f} & -2J^{f} & U^{f}
\end{bmatrix},
\label{Hamitonian_fermi_L2}
\end{equation*}
while  the ground state $H_{B}^{f}$ as the initial state of the battery is given by
\begin{equation}
\rho(0) =\begin{bmatrix}
\frac{1}{4}(1-a) & b & b & \frac{1}{4}(1-a)\\\\
b & \frac{1}{4}(1+a) & \frac{1}{4}(1+a) & b\\\\
b & \frac{1}{4}(1+a) & \frac{1}{4}(1+a) & b\\\\
\frac{1}{4}(1-a) & b & b & \frac{1}{4}(1-a)
\end{bmatrix},
\label{initial_state}
\end{equation}
where $a=\frac{U^f}{\sqrt{16{J^f}^2+{U^f}^2}}$ and $b=\frac{J^{f}}{\sqrt{16{J^f}^2+{U^f}^2}}$. The charging Hamiltonian in the Fock basis  reduces to
\begin{equation}
    H_{c}^{f} = U_{c}^{f}(|1010\rangle \langle 1010| + |0101\rangle  \langle 0101|)
\end{equation}
which is used upto a certain time $t$ to charge  the battery, resulting to an evolved state, 
\begin{equation}
\rho(t) = \begin{bmatrix}
\frac{1}{4}(1-a) & be^{-itU^f_{c}} & be^{-itU^f_{c}} & \frac{1}{4}(1-a)\\\\
be^{itU^f_{c}} & \frac{1}{4}(1+a) & \frac{1}{4}(1+a) & be^{itU^f_{c}}\\\\
be^{itU^f_{c}} & \frac{1}{4}(1+a) & \frac{1}{4}(1+a) & be^{itU^f_{c}}\\\\
\frac{1}{4}(1-a) & be^{-itU^f_{c}} & be^{-itU^f_{c}} & \frac{1}{4}(1-a)
\end{bmatrix}.
\end{equation}
The  work output in this case simplifies as
\begin{equation}
    W^{f}(t)=\frac{{J^f}^2}{{J^f}^2+\frac{(U^{f})^2}{16}} (1-\cos(tU^{f}_{c})).
    \label{work_f}
\end{equation}
\begin{figure*}
     \centering
     \includegraphics[width=0.9\textwidth]{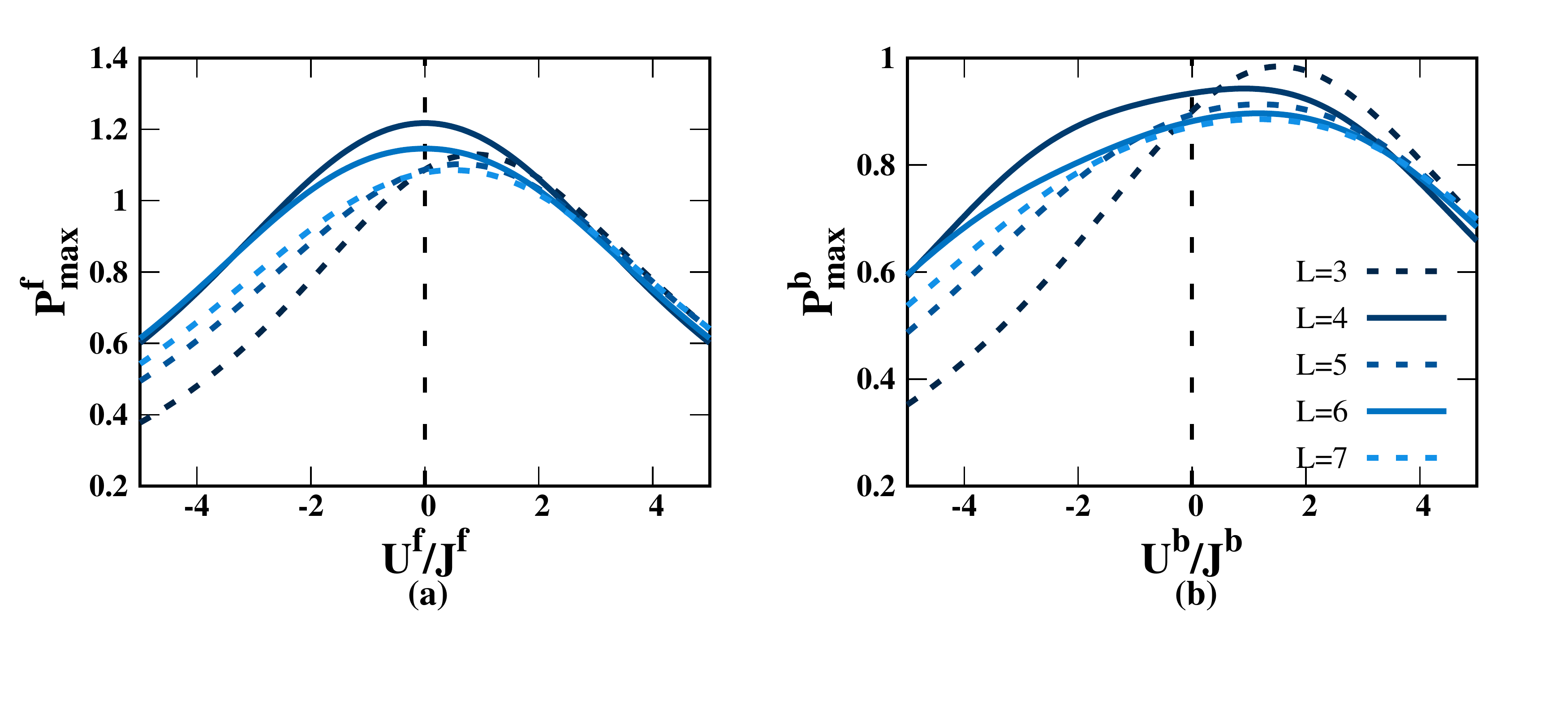}
     \caption{(Color Online.) (a) Power output, $P_{\max}^{f}$, (ordinate) with respect to $U^{f}/J^{f}$ (abscissa) of the battery constructed via  the Fermi-Hubbard model.  (b) $P_{\max}^{b}$  (ordinate)  $U^{b}/J^{b}$ (abscissa) for bosonic systems. The system is half-filled in both the cases and the initial state of the battery is the ground state of the system. In case of Bose-Hubbard model, at most two bosons per sites are only allowed.  The charging of the battery is performed by using on-site interaction with the strength  being $U_{c}^{f} = U_{c}^{b} = 2$. Solid lines correspond to even number of lattice sites, \(L\) while the dashed lines represent the odd \(L\). In both the cases, dark to lighter shades occur with the increase of \(L\).  Both the axes are dimensionless.
     }
     \label{fig:fig1}
 \end{figure*}
Following the same prescription, we also calculate (for detailed calculation, see Appendix-A) the total work output considering BH model for the same time interval $t$,which is given by
\begin{equation}
    W^{b}(t)=\frac{{J^b}^2}{{J^b}^2+\frac{(U^{b})^2}{16}} (1-\cos(tU^{b}_{c})).
    \label{work_b}
\end{equation}
Hence, if $J^{f} = J^{b}$, $U^{f} = U^{b}$ and $U^{f}_{c} = U^{b}_{c}$, the average work output in both the cases are same. 
\end{proof}

\textbf{Remark.} Although  the number of different basis states appears due to exchange symmetry for different spin-statistics to both fermions and bosons, the design of the charging Hamiltonian is responsible for the equal work output obtained in Proposition 1 for two lattice sites. As mentioned before, the charging Hamiltonian only acts on those lattice sites that consists of atleast two particles and 
the number of such states having two particles on a single site occurs equal number of times in both bosonic and femionic batteries which lead to the equal work output from the battery. 
In the succeeding sections, we report unequal power  from bosonic and femionic batteries for large number of lattice sites. 




\subsubsection*{Arbitrary number of lattice sites: Bosons vs. Fermions}

Let us now move further  and 
consider  a lattice having site more than two. First we consider three sites occupied by three particles, bosons or fermions. Unlike the previous case,  we establish a hierarchy between batteries with the performance with the bosons and fermions. 

\textbf{Proposition 2.}  \emph{The battery composed of  three lattice sites filled with three fermions is better in terms of  the work output than that of the bosonic systems in the absence of on-site interaction of the battery Hamiltonian, provided  the charging strength of the on-site interactions for both fermions and bosons are same, i.e.,   $U^{f}_{c} = U^{b}_{c} = U_c\,\, $. }

\begin{proof}
Following the same procedure (see Appendix-B) as in the previous proof,  we calculate $W^{f}(t)$ and $W^{b}(t)$ for a lattice with sites, $L = 3$, occupied with three fermions, and three bosons governed by FH and BH  respectively, having $U^{\mu} = 0$, \(\mu= f, b\). 
If the value of the charging strength for both the cases  are identical,  the difference between the work output turns out to be
\begin{equation}
  W^{f}(t) - W^{b}(t) =  0.13 (1-\cos (t U_{c})),
  \label{eq:workf3}
\end{equation}
which is positive and hence the proof.
Note here that the output work is independent of any system parameters since  the Hamiltonian is normalized and the  spectrum is bounded between \(-1\)  and  \(1\) which allows
us to compare two different models in the same footing. Since \(U^{\mu}=0\),  the expression of the Hamiltonian only involves \(J^{\mu}\) which gets cancelled after  the normalization.
Moreover, the normalization is also responsible for the prefactor \(0.13\) in Eq. (\ref{eq:workf3}) (comparing Eqs. (\ref{eq:workfer3}) and (\ref{eq:workbosons3}).

\end{proof}

 \begin{figure}
     \centering
     \includegraphics[scale=0.35]{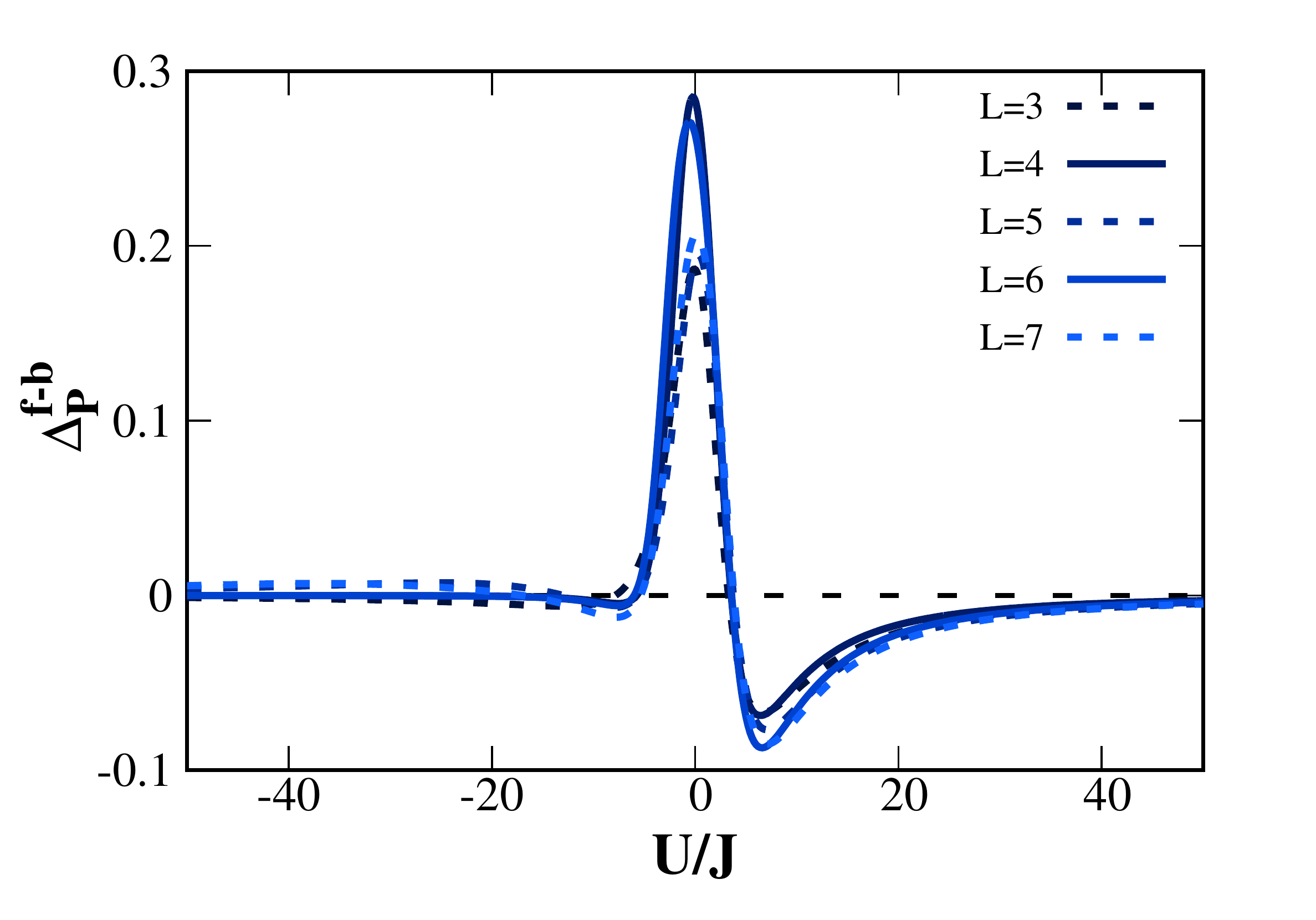}
     \caption{ (Color Online.) \emph{Bosonic vs. fermionic QBs.}  Performance score, $\Delta_P^{f-b} = P_{max}^{f} - P_{max}^{b}$  (vertical axis)  against $U/J = U^b/J^b = U^f/J^f$ (horizontal axis). All other specifications are same as in Fig. \ref{fig:fig1}. Both the axes are dimensionless. 
     }
     \label{fig:fig2}
 \end{figure}
 
 \begin{figure}
     \centering
     \includegraphics[scale=0.35]{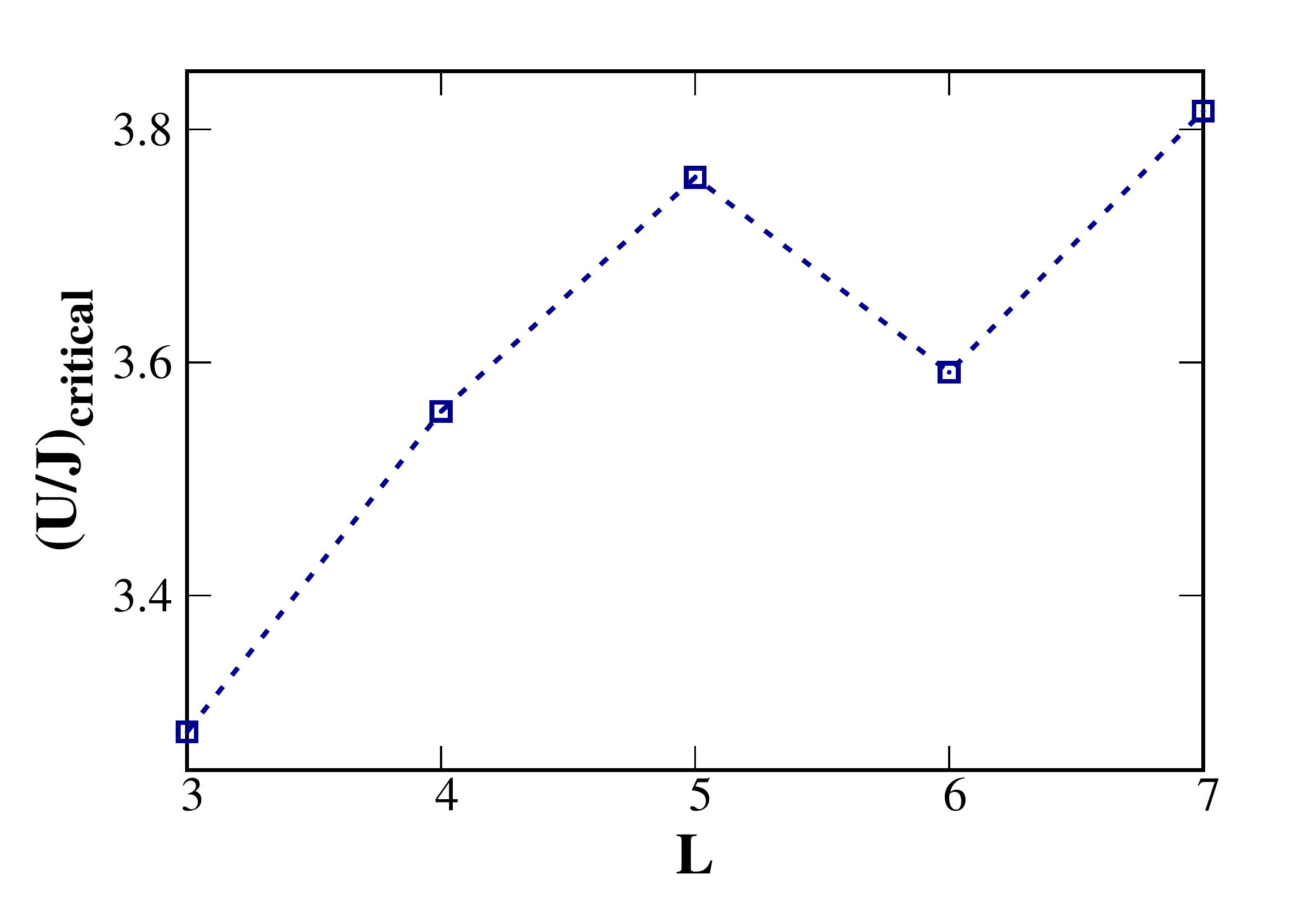}
     \caption{(Color Online.) \emph{Scaling of hierarchy.} The critical value of \(U/J\), denoted by $(U/J)_{critical}$ (\(y\)-axis) above which the batteries build with the BH model can store more energy than that of the  FH  ones with respect to lattice sites $L$ (\(x\)-axis). All the axes are dimensionless.
    }
     \label{fig:fig3}
 \end{figure}

  \begin{figure}
     \centering
     \includegraphics[scale=0.35]{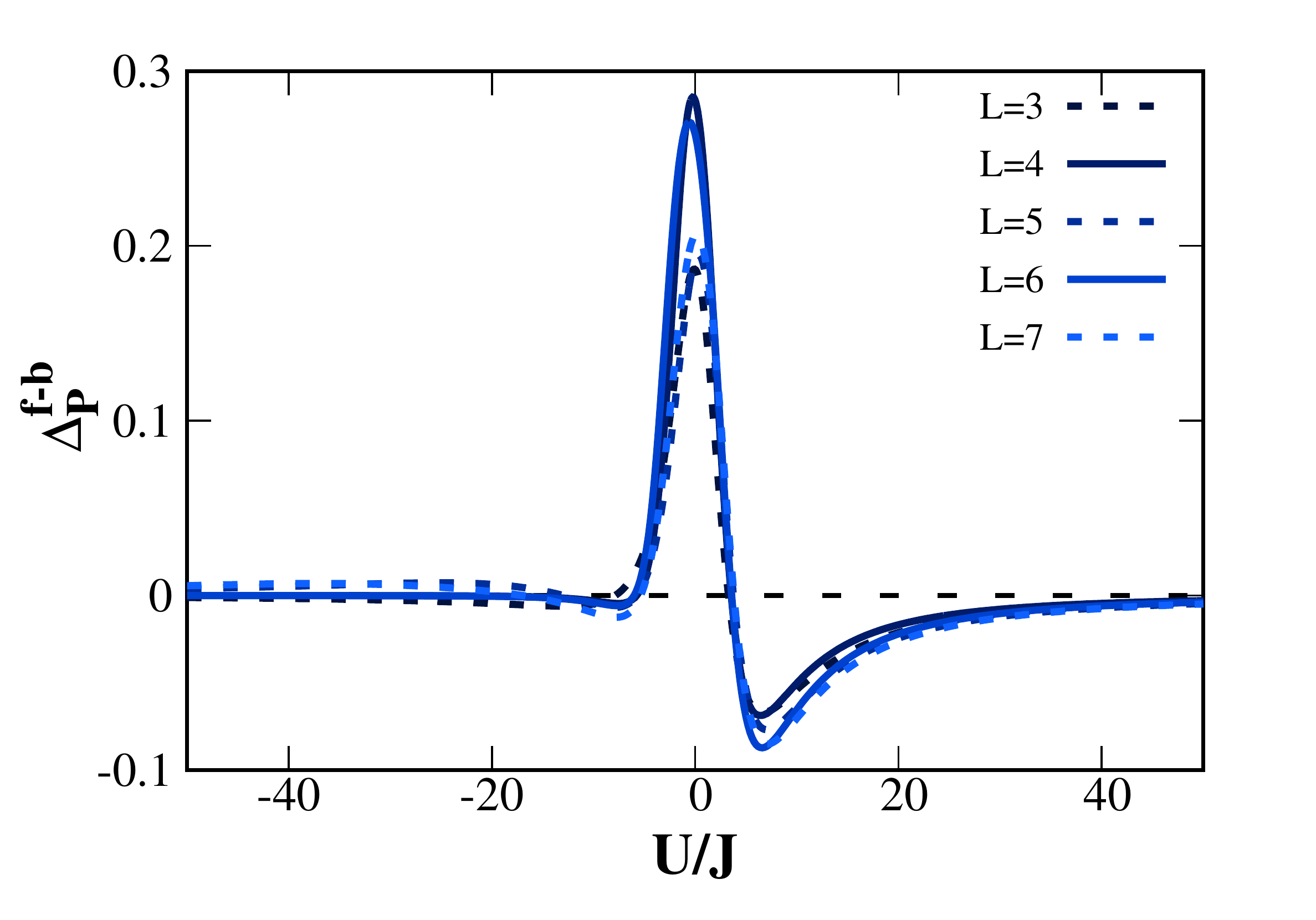}
     \caption{(Color Online.) \emph{Comparison between fermionic and  bosonic QBs with high \(U/J\) values.} $\Delta_P^{f-b} = P_{max}^{f} - P_{max}^{b}$  (ordinate)  vs. $U/J = U^b/J^b = U^f/J^f$ (abscissa). All other specifications are same as in Fig. \ref{fig:fig1}. Note that for high \(U/J\) values, power vanishes in both the cases, thereby obtaining the vanishing performance score.  
     Both the axes are dimensionless.
     }
     \label{fig:high_uj}
 \end{figure}

The Proposition 2 indicates that the  increasing value of lattice sizes and the number of particles can have  significant effects on the power output for these two models. In particular, identifying parameter range where the FH-battery outperforms the  BH ones can be an interesting question to address with \(L\geq 3\). Towards the aim, the initial battery-state is considered to be
 the ground state of the FH lattice  filled up with $N^{f}_{\uparrow} = \lfloor L/2 \rfloor + L (\mbox{mod} \, 2)$ and $N^{f}_{\downarrow} = \lfloor L/2 \rfloor$ fermions where total number of fermions, $L= N^{f} = N^{f}_{\uparrow} + N^{f}_{\downarrow}$ while the  BH-battery is occupied with $N^{b} = L$ number of bosons with $2$ particles per site, i.e., a single site can occupy at most two bosons -- distribution of fermions and bosons in this way is called half-filling. We will lift the restriction of particles per site in the succeeding section. 
 In the rest of the paper, we carry out our analysis of  \(P_{\max}^\mu\) by varying $U^{\mu}/J^{\mu}$,  $(\mu = f,b)$
  since the various phases like Mott-insulator, superfluid, Fermi liquid  and  quantum phase transitions can  successfully be described in the different limits of this ratio. 
 Moreover,   in the entire calculation, we take the strength of the charging field as $U_{c}^{f} = U_{c}^{b} = 2$.  Notice, however, that with the increase of the charging on-site interactions, the power gets enhanced. It can also be understood from the  expressions of work in Eqs. (\ref{work_f}),  (\ref{work_b}) and (\ref{eq:workf3}) which clearly show that the maximum power is obtained for small time when one increases \(U_c^\mu\). 
 %

\emph{Contrasting trends for FH- and BH-batteries.} The patterns of \(P_{\max}^{\mu} \) with  $U^{\mu}/J^{\mu}$ for a paradigmatic example of half-filling of lattice sites both for fermions and bosons are depicted in  Fig. \ref{fig:fig1} and we observe that the contrasting behavior emerges for bosons and fermions -- (1) the FH-based battery produces more power output than that of the BH model almost in the entire range of \(U^{\mu}/J^{\mu}\). We will determine the exact range of advantage obtained via fermionic systems  in Figs. \ref{fig:fig2}, \ref{fig:high_uj}, and \ref{fig:fig3} which we will discuss later; (2) In case of even number of lattice sites with the FH model, \(P_{\max}^f\)  is symmetric  about $U^{f}/J^{f} = 0$-line, thereby leading to maximum average power output with \(U^f =0\), although no such symmetry is observed in case of bosons; (3) In the half-filling regime, among all the lattice sites considered, i.e., when \(3 \leq L \leq 7\),  we find that \(P_{\max}^f\)  reaches its maximal value for \(L=4\) while  \(P_{\max}^b\) shows maximum with \(L=3\) and  $U^{b}/J^{b}>0$. Although, there is, in general, no visible correlation between lattice size and higher work output,  \(P_{\max}^\mu\) converges to a certain value for all values of \(L\) in presence of strong repulsive and attractive interactions, thereby illustrating a site-independent power output. 

  
 \begin{figure*}
     \centering
     \includegraphics[width=\textwidth]{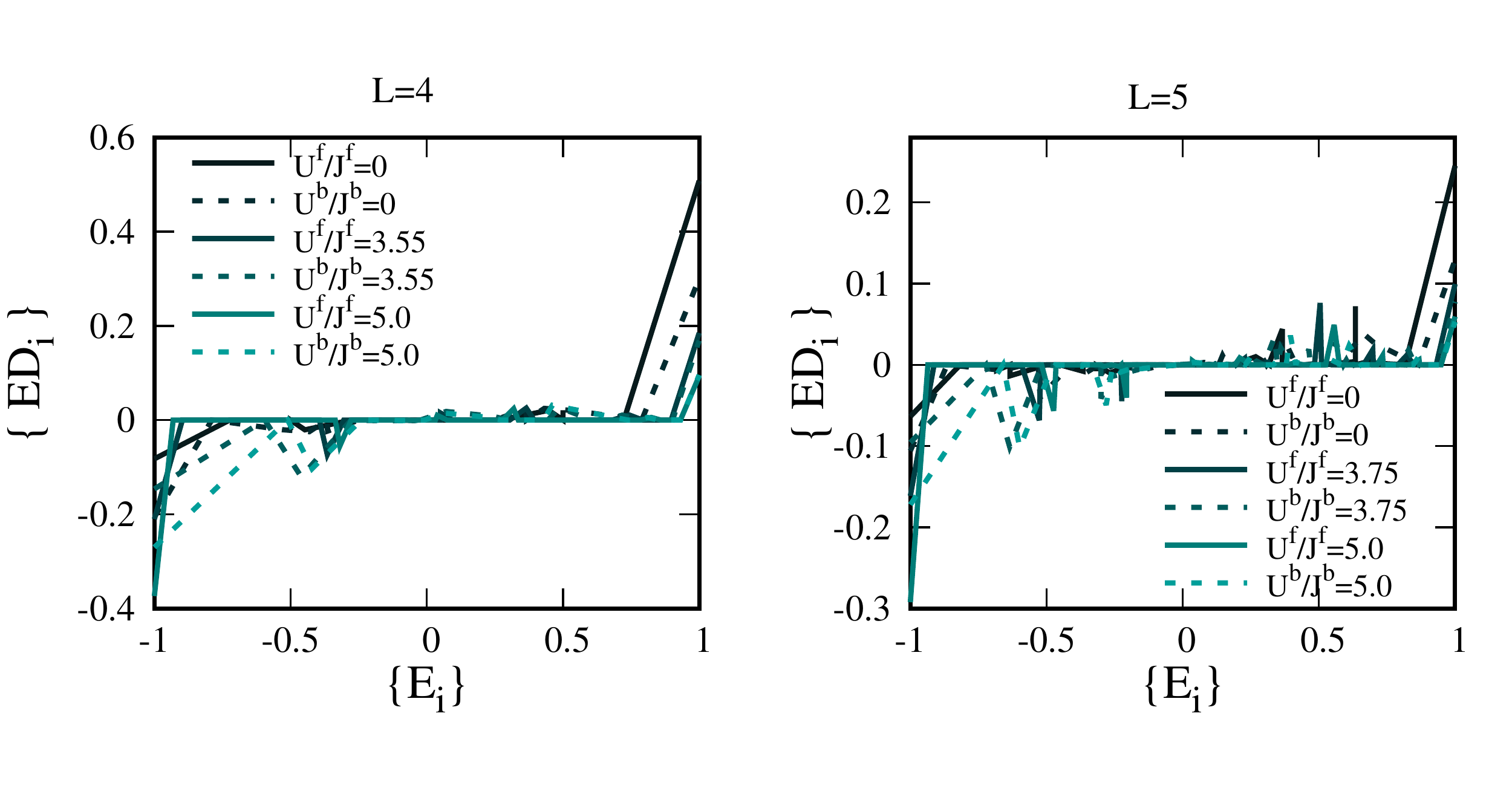}
     \caption{(Color Online.) $\{ED_i\}$ against the corresponding energy eigenvalues of normalized Hamiltonian, $\{E_i\}$ for two different lattice sites, \(L=4\) (left) and \(L=5\) (right).
     Here $U_c=2$. Solid (dashed) lines correspond to fermions (bosons). With the increase of $U^{\mu}/J^{\mu}$, the shades (dark to light) get lighter.
     Both the axes are dimensionless.
     }
     \label{fig:energy_dist}
 \end{figure*}

To compare the batteries constructed with fermionic and bosonic systems, we introduce a quantity which we call the performance score,  
\[\Delta_P^{f-b}= P_{\max}^f - P_{\max}^b
,\]
 by fixing $U^{f} = U^{b} = U$ and $J^{f} = J^{b} = J$. From Fig. \ref{fig:fig2}, we observe that when  $-5\lesssim U/J < 0$, i.e. with attractive on-site interactions, FH-batteries can always store (extract) more energy than that of batteries with BH model  although the situation changes when \(U/J \gtrsim 3\). Specifically, there exists a critical \(U/J\)-value, $(U/J)_{critical}$, above which the bosonic systems can produce more power than that of the fermionic systems, i.e., \(\Delta_P^{f-b} <0\) when \(U/J >(U/J)_{critical}\) . We also notice that \((U/J)_{critical}\) depends on \(L\) as shown in Fig. \ref{fig:fig3} which indicates that with the increase of \(L\), a higher on-site interaction is required to achieve a higher power by using  the BH model than that of the FH one.

If one increases the value of on-site interaction, the system undergoes a phase transition from superfluid phase to the Mott-insulator one, i.e., the probability of hopping between the lattice sites decreases and becomes vanishingly small at some point. Moreover, due to the half-filled scenario,  all the lattice sites then contain only one particle effectively and due to the designing of the charging Hamiltonian, the excitation in the system becomes minimal resulting to a very low power output from the battery. It implies that the difference of the output power between  $\textbf{FH}$ and $\textbf{BH}$ systems also goes to zero (see Fig. \ref{fig:high_uj}) since at a very high value of $U/J$, the power output from the individual system also vanishes.
In other words,  in an extreme scenario, when both charging  and the battery Hamiltonian are in Mott insulating phase, i.e., when \(U >> J\), the battery cannot be charged while moderate values of \(U/J\) is favourable to store the power in the battery as also shown in Fig. \ref{fig:high_uj}.

 \emph{$(U/J)_{critical}$ and the energy distribution. In order to explain the $(U/J)_{critical}$ value, we check the energy distribution of the state that produces maximum average power. More precisely, we calculate the inner product between the evolved state giving maximum power  and eigenstates of the corresponding Hamiltonian, i.e., 
if the state that stores maximum average power is $|\psi^{\mu}(t)\rangle$ (\(\mu = f, b\)),  we compute the quantity \(p_i^\mu = |\langle \epsilon_i^\mu | \psi^\mu (t)\rangle|^2 \) where \(|\epsilon_i^\mu\rangle\) is the eigenstate of the  \textbf{BH} or \textbf{FH} models having energy \(E_i\). To identify the dynamically preferred state,  we compute the  energetically preferred distribution as
\begin{equation}
    \{ED_i^\mu\} = \{p_i^\mu E_i^\mu\} ~~ \forall ~~ i =\{1 \ldots N_E^\mu\}
\end{equation}
where $N_E^\mu$ is the number of eigenvectors, $\{|E_1^\mu\rangle,|E_2^\mu\rangle,\ldots,|E_{N_E}^\mu\rangle\}$ in both Bose and Fermi cases. 
We study $\{ED_i^\mu\}$ against the corresponding $\{E_i\}$ in Fig. \ref{fig:energy_dist}. It shows that  when $U^{\mu}/J^{\mu}$ (\(\mu =f, b\)) is less than the critical point where femionic battery is better than that of the bosonic ones, there is more energetic contribution from the excited state in the Fermi case as compared to the Bosonic ones. After the critical point, the situation reverses and  the excited states in the Bose-Hubbard model contribute more that that of the Fermi-Hubbard case. }

 

\section{Effects of filling factor and temperature  on average power output }
\label{sec:fillingtemp}

Upto now, the entire analysis is carried out by considering the half-filling and when the battery is the ground states of the Hamiltonian.  Let us  lift both the restrictions and study their consequences on the performance of the QB. 
 
First we explore the dependence of \emph{filling factors} on the power output of the battery. 
Before going further, let us first discuss two extreme situations for which the power outputs vanish when the battery is  made of \emph{fermions}.

\textbf{Remark 1.} For a fixed lattice site, if all the lattice sites are completely occupied by up or down or both up and down fermions allowed by the Pauli exclusion principle, no work can be extracted from the system since no excitation is possible in this scenario.

\textbf{Remark 2.} Suppose all the lattice sites are filled with down (up) fermions. If we now increase the number of up (down) fermions one by one on a lattice, the power output again vanishes.  
This is due to the fact that  in this process,  the charging Hamiltonian comes out to be an identity matrix multiplied with a constant, which is the strength of the charging field and after evolving  for a time interval $t$, the evolved state $\rho(t)$ remains  identical with the initial ground state $\rho(0)$. Hence to obtain a non-trivial power output from the QB, the number of up and down fermions in the system of \(L\) lattice sites  must be upper bounded by $L-1$.

In the fermionic system, we also find the following: \\
\textbf{Observation 1.} The \emph{maximum extractable power} is same under the exchange of the total number of up and down fermions in the system i.e., $P_{\max}(N^{1}_{\uparrow},N^{2}_{\downarrow})=P_{\max}(N^{1}_{\downarrow},N^{2}_{\uparrow})$ where \(N^i, \, i =1, 2\) is the number of up (down) and down (up) fermions respectively. Moreover,  we  notice that $P_{\max}(N^{1}_{\uparrow},N^{2}_{\downarrow})=P_{\max}(L-N^{1}_{\uparrow},L-N^{2}_{\downarrow})$. \\

 \begin{figure}[h]
     \centering
     \includegraphics[width=0.5\textwidth]{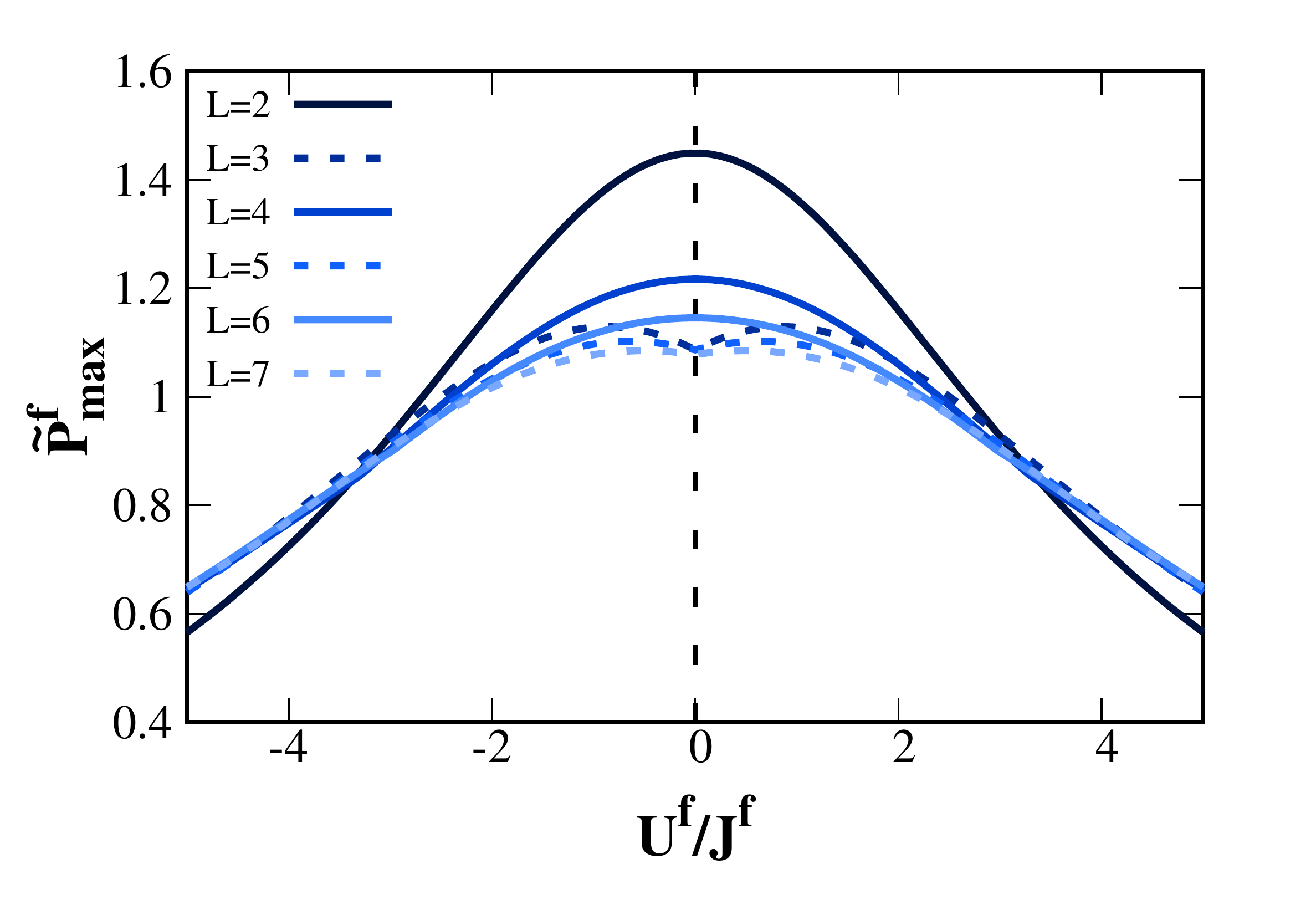}
     \caption{(Color Online.) Variation of  $\widetilde{P}_{\max}^{f}$ (see text for the definition)) (ordinate) vs. $U^{f}/J^{f}$ (abscissa). Notice that the symmetry missing around \(U^f/J^f =0\)-line in  Fig. \ref{fig:fig1}  for odd lattice sites can be attained by considering the quantity $\widetilde{P}_{max}^{f}$ obtained after maximizing over configurations. All other specifications are the same as in Fig. \ref{fig:fig1}. Both the axes are dimensionless. 
     }
     \label{fig:fig4}
 \end{figure}
 
 \begin{figure}
     \centering
     \includegraphics[width=0.5\textwidth]{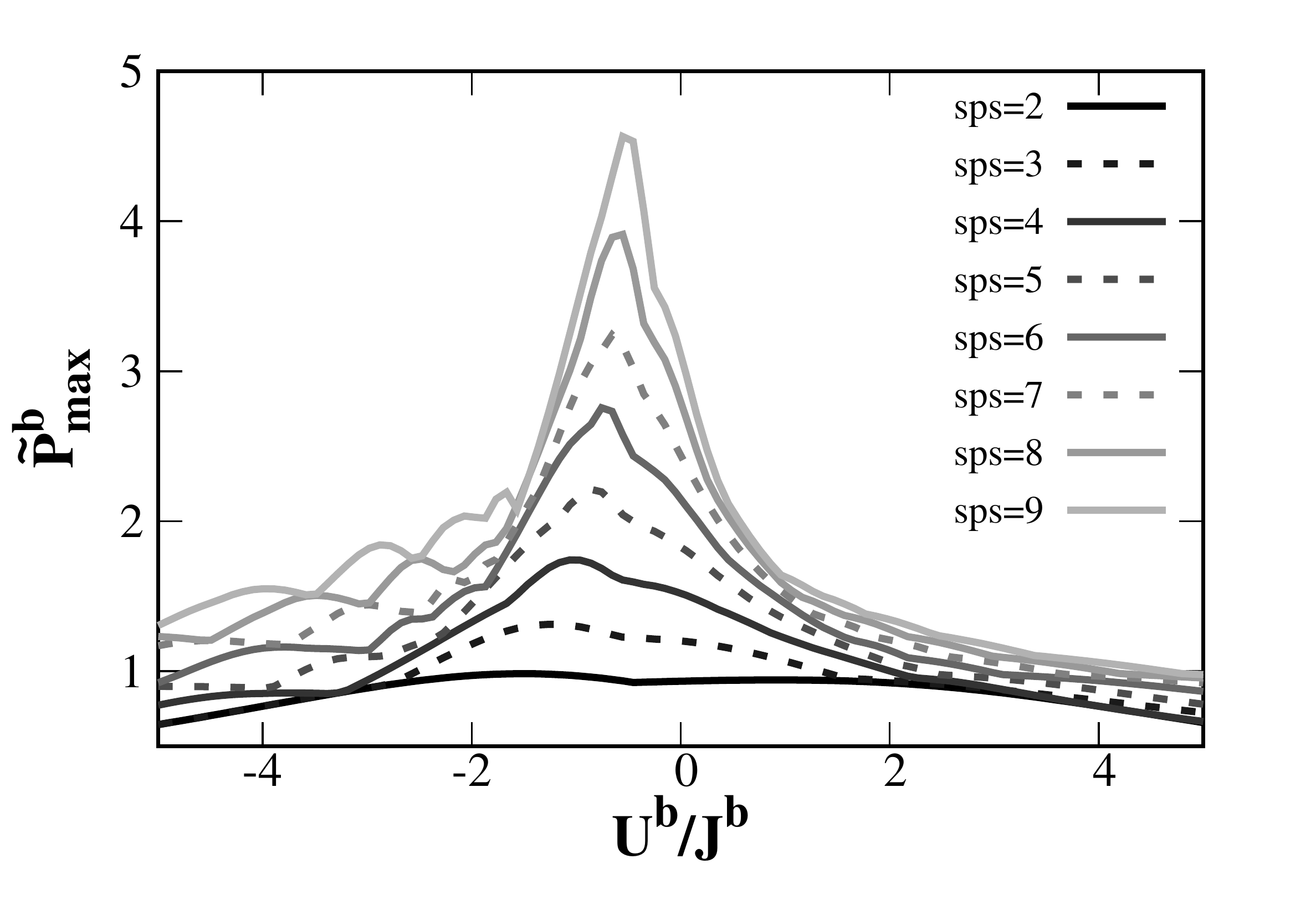}
     \caption{(Color Online.) $\widetilde{P}_{\max}^{b}$ (vertical axis)  vs. $U^{b}/J^{b}$ (horizontal axis) for the BH model-based battery. Solid and dashed lines correspond to even and odd values of \emph{states per site (sps)} respectively which can go at most \(n\). Here \(L=4\). The optimization involved in $\widetilde{P}_{\max}^{b}$ is performed over all the \(nL\) configurations. Dark to lighter shade indicates the increase of state per site. All the axes are dimensionless.
     }
     \label{fig:fig5}
 \end{figure}

Following all these symmetries, let us first calculate the  number of configurations that are giving the non-trivial dynamics of a fermionic system having total number of lattice sites $L$.
The constraints which dictate the configurations, given by 
\begin{enumerate}
    \item $N_{\uparrow}^{1} + N_{\downarrow}^{2} \leq L$, 
    \item $N_{\uparrow}^{1}, N_{\downarrow}^{2} \geq 1$, and
    \item $N_{\uparrow}^{1}, N_{\downarrow}^{2}$ := $L-N_{\uparrow}^{1}, L- N_{\downarrow}^{2}$.
\end{enumerate}
Therefore, the total number of configurations $\mathcal{S}$ that is contributing to the output power is given by
 \begin{equation}
    \mathcal{S} = (L-1) + \{(L-2) - 1\} + \ldots + \{(L - \big\lfloor\frac{L}{2}\big\rfloor) - \big\lfloor \frac{L}{2}\big\rfloor + 1\}.
    \label{sum}
\end{equation}
It is easy to check that for even  $L$, $\mathcal{S} = \frac{L^2}{4}$ while $\mathcal{S} = \frac{L^2-1}{4}$ when $L$ is odd. For example, 
with \(L=2\), the total number of possible configuration is seven. However, incorporating all the aforementioned symmetries, we observe that only a particular configuration among all those choices  are responsible for the maximum amount of  power from the battery,  which turns out to be $N^{f}_{\uparrow} = 1$ and $N^{f}_{\downarrow} = 1$ for the entire parameter regime of $U^{f}/J^{f}$ as counted in Eq. (\ref{sum}).
However,  by increasing the lattice sites, we obtain the maximum power contribution from different filling factors depending on the tuning parameter,  $U^{f}/J^{f}$. To capture it, we introduce a quantity \(\widetilde{P}_{\max}^f = \max P_{\max}^f\) (see Fig. \ref{fig:fig4}) where the maximization is performed over all possible configurations. 
  First of all,   \(\widetilde{P}_{\max}^f\) decreases with the increase of lattice sites although the decrease rate depends on the  even or odd \(L\). 
 Secondly,  the average power output is symmetric about $U^{f}/J^{f} = 0$ (comparing with Fig. \ref{fig:fig2}). Thirdly,  unlike even number of lattice sites,  the value of \(\widetilde{P}_{\max}^f\) is independent of \(L\) at $U^{f}/J^{f} = 0$ for odd number of lattice sites, although the maxima occurs at some point with \(U^f/J^f >0\) and \(U^f/J^f <0\) symmetrically.  \\

In the case of BH model,  we consider a scenario  where the number of lattice site is fixed to $L$, and particles per site available is at most $n$. Again we examine \(\widetilde{P}_{\max}^b = \max P_{\max}^b\) where the maximization is taken over all the allowed configurations possible under the constraint of \(n\) particles per site, thereby optimizing over  \(n L\) configurations (see Fig. \ref{fig:fig5} for \(L=4\)). For a fixed number of lattice sites, \(\widetilde{P}_{\max}^b\) increases with the increase of \(n\). 
 In contrast to the fermionic battery,  the power output for the bosonic battery is not symmetric about $U^{b}/J^{b} = 0$-line.

 \begin{figure}
     \centering
     \includegraphics[width=8.5cm,height=6.5cm]{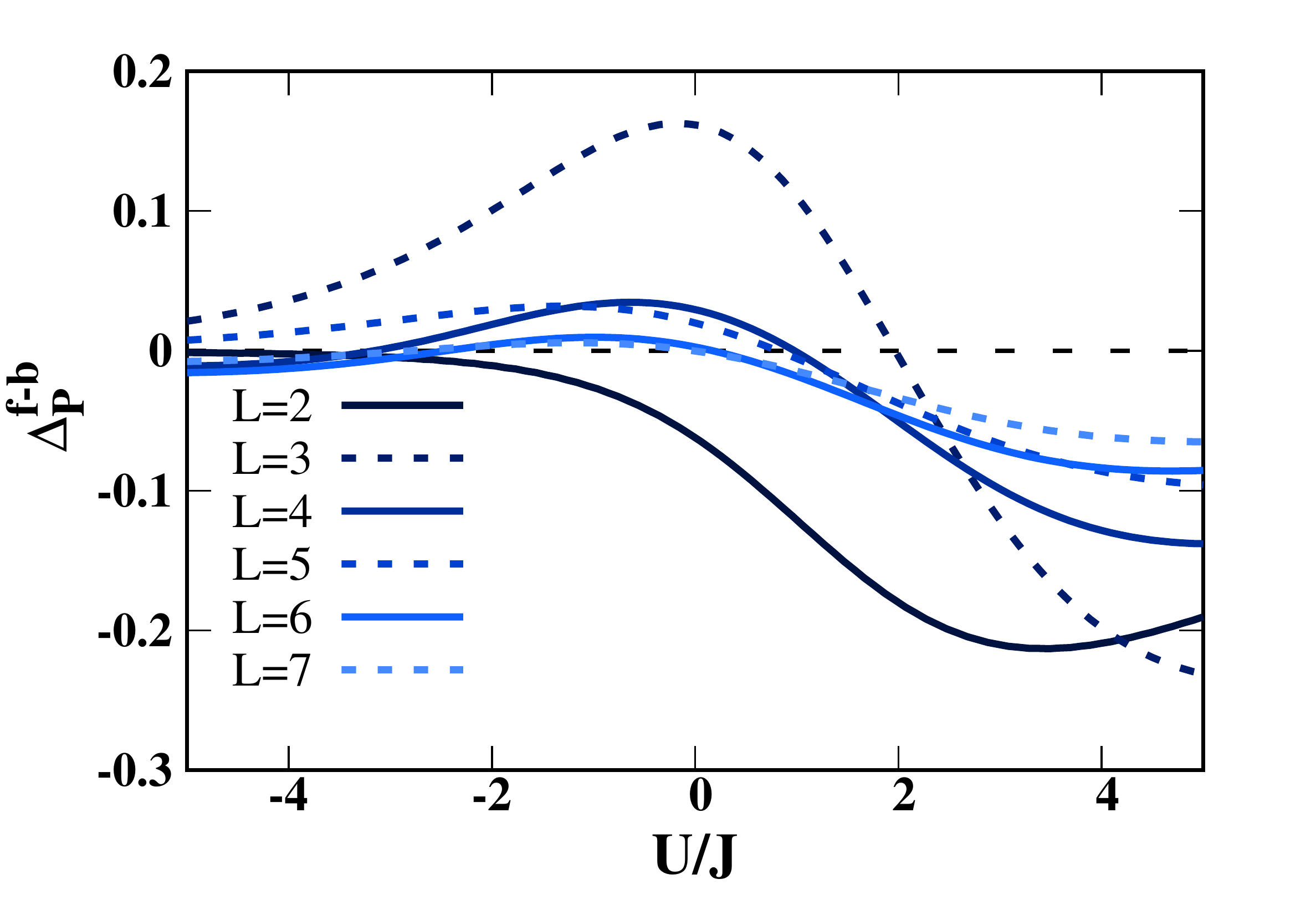}
     \caption{(Color Online.) \(\Delta_P^{f-b}\) (vertical axis) with  $U/J$ (horizontal axis). Here $\beta = 3$. 
     All other specifications are same as in Fig. \ref{fig:fig1}. All the axes are dimensionless.
     }
     \label{fig:fig6}
 \end{figure} 
 

\emph{Role-reversal of bosonic and fermionic batteries depending on temperature.} In a  more realistic situation, one expects that the initial state of the quantum battery is the thermal state or the canonical equilibrium state ($\rho_{th}$) of the Hamiltonian. To illustrate the effects of temperature on the maximum  average power output of the battery built by the BH and the FH models, we examine the performance score, \(\Delta_P^{f-b}\), by varying $U/J$, where $U^{b} = U^{f} = U$ and $J^{f} = J^{b} =J$ and set $\beta = |J|\beta^{'}$. With the increase of temperature, we find that  Proposition 1  for two lattice sites does not remain valid, i.e., \(P_{\max}^f \neq P_{\max}^b\) with some moderate temperature. Specifically, we find that \(U/J=5\), \(\Delta_P^{f-b}\) becomes negative when the initial state is prepared at \(\beta \lesssim 39.5\), thereby showing that bosonic batteries outperform the fermionic ones. 
Such an advantageous role of bosonic systems persists also for a higher number of lattice sites with a certain \(\beta\) value  and a wide range of \(U/J\) as depicted in Fig. \ref{fig:fig6}.  Specifically, if the initial state is prepared at a very high temperature, the  maximum average power output obtained from the BH models is higher than that of the FH ones in most of the repulsive on-site interaction, i.e., for positive values of \(U/J\).

\section{Robustness of batteries based on Hubbard models in presence of Disorder }
\label{sec:robust}

\begin{figure*}
     \centering
     \includegraphics[width=0.7\textwidth]{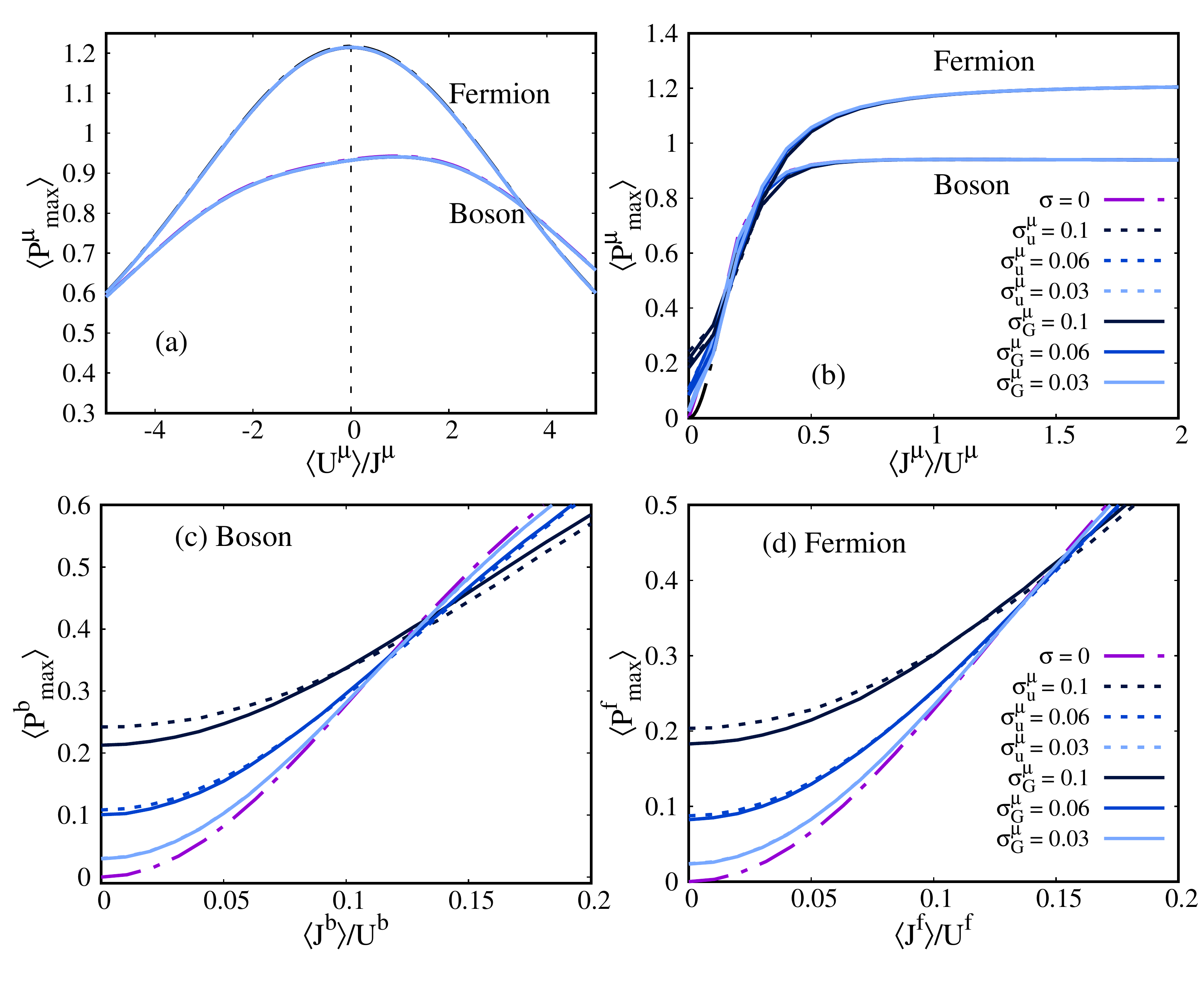}
     \caption{(Color Online.) \emph{Disorder enhanced power.} (a)-(b) Quenched averaged power, $\langle P_{max}^\mu \rangle $, by varying  $\langle U^\mu\rangle/J$ and $\langle J^\mu \rangle/U$ with \(\mu =f, b\). Here \(L=4\) and the initial state is the ground state of the system.  \(\sigma^\mu_G\) and \(\sigma_u^\mu\) represent  the standard deviations of the  Gaussian (solid line) and the uniform  (dashed line) distributions respectively from which the  on-site interactions and the hopping are randomly chosen. Note that \(\sigma=0\) represents the ordered systems (dashed-dotted line). The higher $\langle P_{max}^\mu \rangle $ value correspond to the disordered FH models while the lower values are for the bosonic systems. 
 (c)-(d) $\langle P_{max}^b \rangle $ and $\langle P_{max}^f\rangle $   are plotted with respect to  $\langle J^b \rangle/U$ and  $\langle J^f \rangle/U$  respectively. Dark to lighter shade indicates the decrease of standard deviation, except for the ordered case. In both the situations,  systems with impurities turn out to be  a better storage device than that of the ordered systems, thereby showing disorder induced power.  
 All the axes are dimensionless. 
     }
     \label{fig:Disorder}
 \end{figure*}

In an ultracold atom experiment, disorder can be introduced in the system in a controlled manner \cite{aditireview}, leading to the quenched disordered system. In particular,   van der Waals losses are mitigated by placing atoms at a significant distance from the atom chip, which results in an adjustment of the magnetic wire, leading to the uniform disorder in the on-site intraparticle interactions, \(U^\mu/J^\mu\) (\(\mu=f, b\))  \cite{Uniform_Disorder2, Uniform_disorder1} of the Hubbard Hamiltonian.  The disorder is chosen from uniform distribution,  $U^\mu_u/J^\mu \in [a,b]$ with mean $\langle U_u^\mu\rangle/J^\mu = \frac{a+b}{2}$ and standard deviation as $\sigma_u^\mu = \sqrt{\frac{(b-a)^2}{12}}$. On the other hand, 
 the local potential can also be sampled from Gaussian distribution \cite{DisorderJ_gaussian},
  $U^{\mu}_G/J \in \frac{1}{\sigma^{\mu}_G \sqrt{2 \pi}} e^{-\frac{1}{2}\left(\frac{x- \langle U_G^\mu\rangle/J^\mu}{\sigma^\mu_G}\right)^{2}}$ with mean \(\langle U_G^\mu\rangle/J^\mu\) and standard deviation \(\sigma^\mu_G\). Here the subscripts, \(u\) and \(G\) in mean and standard deviations represent the uniform and the Gaussian distributions respectively.

By incorporating uniform as well as Gaussian randomness in the on-site interactions, $\langle U_u^\mu\rangle/J^\mu$, and  $\langle U_G^\mu\rangle/J^\mu$ of the batteries built by FH and BH models, we examine the quenched  maximum average power \cite{disorder},  $\langle P_{max}^\mu \rangle$. It is obtained by computing  $P_{max}^\mu$ for every value of $U^\mu/J^\mu$  chosen randomly from both uniform and Gaussian distributions with the corresponding means and the standard deviations. 
The number of realizations considered here for calculations is $2 \times 10^5$  leading to a convergence upto two decimal places. 
 In order to maintain a fair comparison between systems with and without disorder,  we choose \(L=4\) sites with half-filling, the particles per site for bosons are restricted to $2$ while the number of spin-up and spin-down fermions are same. We report that both for bosons and fermions,  $\langle P_{max}^\mu \rangle$ does not change substantially in presence of impurities in the on-site interactions as shown in Fig. \ref{fig:Disorder} (a), thereby illustrating robustness in the performance of the battery against disorder.  
 
  
On the other hand, disorder in the hopping parameter of the Hubbard Hamiltonian can  be realized by modulating the applied electric field of the laser or by doping impurities in the system \cite{DisorderInJ_Nandini, DisorderJ_gaussian}. In this scenario, the quenched disorder power output, \(\langle P_{\max}^\mu \rangle\) from the FH and the BH models are again computed by varying  \(\langle J^\mu_{G (u)}\rangle/U^\mu\)  for different but fixed standard deviations. 
%
%
 Like randomness in the on-site interactions, when both uniform and Gaussian disorders are impinged in the hopping terms of the Bose- and Fermi-Hubbard Hamiltonians, thereby changing the initial state of the battery,   no significant consequences on power of the battery are observed over the ordered case (see Fig. \ref{fig:Disorder} (b)).  Interestingly, however, when inspected closely, we notice that for \(\langle J^{\mu}_G\rangle/U^\mu <0.15\) (\(\langle J^{\mu}_u\rangle/U^\mu <0.15\)), the quenched averaged power output from the disordered battery is higher than that of the ordered case, thereby showing improvements in the performance of the battery in presence  of impurities in hopping. Such a \emph{disorder enhanced  power}  is discovered both for fermionic as well as bosonic systems. 
Output power is a monotonically increasing function of $J^{\mu}$, i.e., in the ordered scenario with $\sigma = 0$ in  Figs. 10(c) and (d), the output power is a monotonically increasing function of $J^\mu$, i.e., $J^{\mu} (\epsilon) > J^{\mu} (\epsilon^{'})$ where $\epsilon > \epsilon^{'}$ in some parameter regime. Moreover, it has a very high slope in the same parameter regime,  which implies that \(\langle P^{\mu}_{\max} \rangle (\langle J^{\mu} \rangle_{i+1}) - \langle P^{\mu}_{\max} \rangle (\langle J^{\mu} \rangle_{i}) >> \langle J^{\mu} \rangle_{i+1}- \langle J^{\mu} \rangle_{i}\) around $J^{\mu} = \epsilon$ with $\epsilon \to 0$.
 By introducing disorder into the system, we perform averaging over different realizations of $J^{\mu}$ from a range of $\eta + 3\sigma$ to \(\eta - 3\sigma\)  in the parameter space, where $\eta$ is the mean value of $J^{\mu}$ and $\sigma$ is the standard deviation. Since there are parameter values of \(J^{\mu}\) where the slope of the power with $\sigma = 0$ is much higher than unity,  the power obtained between $[\eta, \eta + 3\sigma]$ is  much higher than that obtained in the region with $[\eta - 3\sigma,\eta]$, thereby providing higher values in power for the disordered case than that of the ordered ones.
 Specifically, advantage in power with impurities is detected in the regime \(0\leq \langle J^{b}_u\rangle/U \leq 0.12\) for bosons while the range of parameters, \(\langle J^{f}_u\rangle/U \in [0, 0.148]\), increase in case of FH model (comparing Figs. \ref{fig:Disorder} (c) and \ref{fig:Disorder} (d)). Note, interestingly, that the power obtained in the disordered case is almost equal to the battery with ordered Hamiltonian in the entire range of parameters, thereby establishing the robustness against impurities in such design of a battery. 
 

\section{Conclusion}
\label{sec:conclu}

Batteries are  integral parts of any technology for storing power and utilizing it as a source of energy at any point in time. We know that the existing battery that we termed as a classical batteries convert the chemical energy to the electrical one and is quite useful although current technological developments demand miniaturization, which inevitably has a possibility to enter the quantum regime. To fulfil the requirements, quantum technologies are designed which also necessitates the modelling of the storage device based on quantum mechanics, leading to quantum batteries (QB). In recent times, several experimental proposals for QBs using quantum dots coupled to cavities, superconducting qubits have been developed and realized.
					
In the current work, we designed a quantum battery in presence  and absence of impurities using ultracold atoms in optical lattices which can be implemented via currently available technologies. In particular, we prepared an initial state of the quantum battery as the ground or the thermal state of the Fermi- Hubbard as well as Bose-Hubbard models. The charging process of the battery is carried out by tuning the on-site interactions. We showed that in the case of lattice sites more than two, and with half-filling, the QB based on the Fermi-Hubbard model can store a higher amount of energy compared to the battery with the Bose-Hubbard model provided the on-site interactions are attractive or repulsive with moderate values. The situation gets reversed if the temperature in the initial state is reasonably high. Moreover, we noticed that the filling factor in both the bosonic and fermionic models plays a crucial role in the power output of the battery. Specifically, the maximum average power after optimizing over all the configurations increases with the increase of particles per site in the case of bosons.

One of the success stories in the ultracold atomic systems is the realization of disorder in a controlled manner. We found that the randomness chosen from the uniform and the Gaussian distributions  in the hopping and in the on-site interactions does not affect the performance of the QB significantly, thereby demonstrating the advantage of preparing these batteries based on ultracold atoms. We also identified a region of mean hoping strength below which the quenched averaged power is higher for the disordered system than that of the ordered ones -- \emph{disorder enhanced power}. The entire engineering of QB proposed via bosonic and fermionic systems opens up a possibility to design thermal machines based on Hubbard models, realizable in laboratories, and at the same time, it can pinpoint the regime in which the machinery based on bosons is better than that of fermions and vice-versa.

 
\section{acknowledgements}
The authors acknowledge the support from the Interdisciplinary Cyber Physical Systems (ICPS) program of the Department of Science and Technology (DST), India, Grant No.: DST/ICPS/QuST/Theme- 1/2019/23. We  acknowledge the use of a modern PYTHON library for general purpose condensed matter physics \cite{quspin,quspin1}, and the cluster computing facility at the Harish-Chandra Research Institute. 

\begin{appendix}

\section{Two-site two-particle system}
Let us consider the scenario when the lattice has two sites occupied with two bosonic particles. The normalized Hamiltonian in the Fock state basis looks like
\begin{equation}
H_{B}^{b} = \frac{1}{\sqrt{16{J^b}^2+{U^b}^2}}\begin{bmatrix}
U^{b} & -2\sqrt{2}J^b & 0\\\\
-2\sqrt{2}J^b & -U^b & -2\sqrt{2}J^b\\\\
0 & -2\sqrt{2}J^b & U^b\\\\
\end{bmatrix}.
\end{equation}
The initial state $\rho(0)$ of the system is the ground state of this Hamiltonian $H_{B}^{b}$, given by
\begin{equation}
\rho(0) = \begin{bmatrix}
\frac{1}{4}(1-a) & \sqrt{2}b^{'} & \frac{1}{4}(1-a)\\\\
\sqrt{2}b^{'} & \frac{1}{2}(1+a) & \sqrt{2}b^{'}\\\\
\frac{1}{4}(1-a) & \sqrt{2}b^{'} & \frac{1}{4}(1-a)\\\\
\end{bmatrix},
\end{equation}
where $a=\frac{U^b}{\sqrt{16{J^b}^2+{U^b}^2}}$ and $b^{'}=\frac{J^{b}}{\sqrt{16{J^b}^2+{U^b}^2}}$. 
We construct the charging Hamiltonian by putting $J^{b} = 0$ which reads as
\begin{equation}
H_{c}^{b} =\begin{bmatrix}
 U^b_{c} & 0 & 0\\
0 & 0 & 0\\
0 & 0 & U^b_{c}\\
\end{bmatrix}.
\end{equation}
After evolving the state $\rho(0)$ by the unitary operator $\mathcal{U}_c = \exp(-i H_{c}^{b} t)$ for a time interval $t$, the resultant state $\rho(t)$ becomes
\begin{equation}
\begin{bmatrix}
\frac{1}{4}(1-a) & \sqrt{2}b^{'}e^{-itU^b_{c}} & \frac{1}{4}(1-a)\\\\
\sqrt{2}b^{'}e^{itU^b_{c}} & \frac{1}{2}(1+a) & \sqrt{2}b^{'}e^{itU^b_{c}}\\\\
\frac{1}{4}(1-a) & \sqrt{2}b^{'}e^{-itU^b_{c}} & \frac{1}{4}(1-a)\\\\
\end{bmatrix}.
\end{equation}
The average work output can then be computed as
\begin{equation}
 W^{b}(t)=\frac{{J^b}^2}{{J^b}^2+(0.25U^{b})^2}(1-\cos(tU^b_{c}))
\end{equation}

\section{Three sites and three particle scenario in absence of on-site interaction }

Let us consider the lattice having three sites and for FH model, it has $N_{\uparrow}^{f} = 2$ and $N_{\downarrow}^{f} = 1$. In absence of $U^{f}$, the three-particle Hamiltonian reads as
\begin{widetext}
\begin{equation}
H_{B}^{f} =  \frac{1}{4\sqrt{2}J^{f}} \begin{bmatrix}
    0 & -2J^{f} & 0 & -2J^{f} & 0 & 0 & 0 & 0 & 0 \\
    -2J^{f} & 0 & -2J^{f} & 0 & -2J^{f} & 0 & 0 & 0 & 0 \\
    0 & -2J^{f} & 0 & 0 & 0 & -2J^{f} & 0 & 0 & 0 \\
    -2J^{f} & 0 & 0 & 0 & -2J^{f} & 0 & -2J^{f} & 0 & 0 \\
    0 & -2J^{f} & 0 & -2J^{f} & 0 & -2J^{f} & 0 & -2J^{f} & 0 \\
    0 & 0 & -2J^{f} & 0 & -2J^{f} & 0 & 0 & 0 & -2J^{f} \\
    0 & 0 & 0 & -2J^{f} & 0 & 0 & 0 & -2J^{f} & 0 \\
    0 & 0 & 0 & 0 & -2J^{f} & 0 & -2J^{f} & 0 & -2J^{f} \\
    0 & 0 & 0 & 0 & 0 & -2J^{f} & 0 & -2J^{f} & 0 \\
    \end{bmatrix}.
    \label{three_sites_H}
\end{equation}
\end{widetext}
Following the same construction procedure as for two lattice sites,  the charging Hamiltonian takes the form as
$$H_{c}^{f}=diag\{U^f_{c},U^f_{c},0,U^f_{c},0,U^f_{c},0,U^f_{c},U^f_{c}\}$$
which leads to the
 average work for the system composed of fermions as
\begin{equation}
  W^{f}(t)=0.75(1-\cos(tU_{c}^{f})).
  \label{eq:workfer3}
\end{equation}

On the other hand, for the BH system with $L = 3$ and $N^{b} = 3$ with maximum two particles per site, the Hamiltonian becomes
\begin{widetext}
\begin{equation}
 H_{B}^{b} =  \frac{1}{(3 + \sqrt{17}) J^b} \begin{bmatrix}
    0 & -2J^b & -4J^b & 0 & 0 & 0 & 0 \\
    -2J^b & 0 & 0 & -2\sqrt{2}J^b & 0 & 0 & 0 \\
    -4J^b & 0 & 0 & -2\sqrt{2}J^b & 0 & 0 & 0 \\
    0 & -2\sqrt{2}J^b & -2\sqrt{2}J^b & 0 & -2\sqrt{2}J^b & -2\sqrt{2}J^b & 0 \\
    0 & 0 & 0 & -2\sqrt{2}J^b & 0 & 0 & -2J^b \\
    0 & 0 & 0 & -2\sqrt{2}J^b & 0 & 0 & -4J^b \\
    0 & 0 & 0 & 0 & -2J^b & -4J^b & 0 \\
    \end{bmatrix}.
\end{equation}
\end{widetext}
 In this case, the charging Hamiltonian reads as
$$H_{c}^{b}=diag\{U^b_{c},U^b_{c},U^b_{c},0,0,U^b_{c},U^b_{c},U^b_{c}\},$$
and the average work turns out to be
\begin{equation}
   W^{b}(t)=0.621(1-\cos(tU_{c}^{b})). 
   \label{eq:workbosons3}
\end{equation}
\end{appendix}

 \bibliographystyle{apsrev4-2}
\bibliography{reference1.bib}

\begin{thebibliography}{76}
\bibitem [{dis()}]{disorder}%
  \BibitemOpen
  \bibinfo {note} {A quenched averaged quantity of any physical observable,
  $M$, calculated over a random distribution $\{Y_j\}$ with a given mean
  $\langle Y \rangle$ and standard deviation, $\sigma_Y$ is calculated as
  \begin{equation} \left\langle M\left(\langle Y \rangle,
  \sigma_{Y}\right)\right\rangle=\iint \cdots \int M\left\{Y_{j}\right\}
  d\left\{Y_{j}\right\} \nonumber \end{equation}}\BibitemShut {NoStop}%
\end{thebibliography}
\end{document}